# DO SPIRALS AND ELLIPTICALS TRACE THE SAME VELOCITY FIELD?

**Tsafrir Kolatt and Avishai Dekel**
Racah Institute of Physics, The Hebrew University
Jerusalem 91904, Israel

tsafrir@astro.huji.ac.il    dekel@astro.huji.ac.il

## ABSTRACT

We test the hypothesis that the velocity field derived from Tully-Fisher measurements of spiral galaxies, and that derived independently from $D_n$-$\sigma$ measurements of ellipticals and S0s, are noisy versions of the same underlying velocity field. The radial velocity fields are derived using tensor Gaussian smoothing of radius 1200 km s$^{-1}$. They are compared at grid points near which the sampling by both types of galaxies is proper. This requirement defines a volume of $\simeq$ (50 h$^{-1}$Mpc)$^3$, containing $\sim$ 10 independent subvolumes, mostly limited by the available ellipticals. The two fields are compared using a correlation statistic, whose distribution is determined via Monte-Carlo simulations. We find that the data is *consistent* with the hypothesis, at the 10% level. We demonstrate that the failure to reject the correlation is not just a result of the errors being big, by using the same method to rule out complete independence between the fields at the 99.8% level. The zero points of the two distance indicators are matched by maximizing the correlation between the two velocity fields. There is a marginal hint that the ellipticals tend to stream slower than the spirals by $\simeq$ 8%. The correlation reinforced here is consistent with the common working hypotheses that (a) the derived large-scale velocity field is real, (b) it has a gravitational origin, and (c) the large-scale velocities of spirals and ellipticals are hardly biased relative to each other. On the other hand, it does not rule out any alternative to gravity where objects of all types obtain similar large-scale velocities.

*Subject Headings:* cosmology: observation — cosmology: theory — large-scale structure of the Universe — galaxies: clustering.

# 1. INTRODUCTION

The techniques developed for measuring distances to galaxies independently of redshifts allow the mapping of the large-scale peculiar velocity field, revealing coherent streaming motions and large divergences. Under the assumption that these velocities were generated by gravity, this development opened a new field of large-scale dynamics, with far-reaching implications on cosmology. For example, the coherent flow constrains the statistical properties of the initial density fluctuations with implications on the nature of the dark-matter constituents (*e.g.* Efstathiou, Bond & White 1993; Kolatt & Dekel 1993; Seljak and Bertschinger 1993), and the local divergences indicate a large value for the cosmological density parameter $\Omega$ (*e.g.* Dekel *et al.* 1993; Nusser and Dekel 1993; Bernardeau *et al.* 1994; Dekel & Rees 1994).

The great impact of these results calls for every possible caution concerning the validity of the interpretation of the measurements as indicating a peculiar velocity field associated with gravity. There is no way to prove that this hypothesis is true, but we should test every possibly-relevant observation which might violate the predictions of the theory, trying our best to rule out the hypothesis. If these are indeed true velocities generated by gravity from fluctuations in the gravitational potential, then, following Galileo, the smoothed velocity field of all test bodies should reflect the same underlying velocity field, independently of their mass. In particular, galaxies of different types should trace the same velocity field after being properly smoothed in the same way over a large scale. The currently available sample of $\sim 3000$ galaxy velocities allows a meaningful consistency check of this hypothesis, by comparing the velocity fields derived independently from spiral and elliptical galaxies at the same positions in space.

The distance indicators arise from the Tully-Fisher (TF) relation for spiral galaxies (Aaronson *et al.* 1982; hereafter S galaxies) and the $D_n$-$\sigma$ relation for elliptical and S0 galaxies (Lynden-Bell *et al.* 1988; hereafter E galaxies). These are empirical intrinsic relations between distance-independent quantities (the rotation velocity in spirals and the dispersion velocity in ellipticals) and distance-dependent quantities (the luminosity in spirals and a "diameter" in ellipticals). The intrinsic relations, up to a normalization zero point, are derived from cluster galaxies that are assumed to be at a common distance for each cluster. When applied to other galaxies, the relation provides estimated relative distances, with an uncertainty of 15% and 21% for Ss and Es respectively.

One worry that has been raised is that the indicated streaming motions are mostly a reflection of systematic environmental variations in the galaxy properties (*e.g.* Silk 1989; Djorgovski, de Carvalho & Han 1989). Efforts to detect correlations between peculiar velocities of each type and certain galaxy characteristics or environmental properties led to null or marginal detections (Burstein *et al.* 1989; Burstein, Faber & Dressler 1990; Burstein 1990). Still, if such correlations do exist, the environmental effects on the different quantities that enter the two different distance estimators may very well be different. This is particularly possible since ellipticals tend to reside in cluster cores while spirals tend to avoid them. Furthermore, while the three-parametered $D_n$-$\sigma$ relation is a straightforward reflection of the virial theorem coupled with a smoothly-varying $M/L$ (Faber *et al.* 1987), the two-parametered TF relation requires an additional physical constraint which is probably imposed at galaxy formation (Gunn 1989). This is a fundamental difference between the two distance indicators, which makes the comparison of the velocity fields derived from Ss and Es very relevant to the question of systematic environmental effects; a strong correlation between the fields would make this possibility less plausible.

Earlier, qualitative comparisons of the radial peculiar velocities of Es and Ss, in certain regions of field and cluster galaxies, indicated a general agreement (Burstein, Faber & Dressler 1990). Our goal is to pursue a quantitative comparison, using an improved analysis and better data.

The data provides noisy radial velocities at the positions of the galaxies. In order to carry out a comparison at the same points in space we need to interpolate the radial velocities of each type into a smoothed radial velocity field at the points of a uniform spatial grid. The smoothing should take into account the variation of the radial direction within the smoothing window, the different non-uniform sampling by each of the two data sets, and the distance measurement errors. For this purpose we use an improved version of the Gaussian tensor window smoothing of *POTENT* [Bertschinger and Dekel 1989 (BD); Dekel *et al.* 1990 (DBF); Bertschinger *et al.* 1990 (BDFDB); Kolatt *et al.* 1994; Dekel *et al.* 1994], which serves there as a first step towards recovering the three-dimensional velocity field using the gravitational ansatz of potential flow, and then revealing the associated mass-density fluctuation field.

A preliminary test in a similar direction has been attempted by Bertschinger (1990), using an earlier data set with only 429 nearby S and 544 Es (Burstein 1990). He found that the velocities traced by the two galaxy types differ systematically, possibly because of a zero-point shift in the nearby spiral data used there. With the extended and carefully-calibrated spiral data now available, and certain improvements in the bias-correction procedure, there is a hope for more conclusive results.

The aim of this paper is thus to test whether the smoothed velocity fields derived independently from spirals and from ellipticals are consistent with being both noisy versions of the same true underlying velocity field. To complement this test, we will check how well can the same method reject the opposite hypothesis, that the fields are independent of each other. In the course of this comparison, we will determine the relative zero-points of the two data sets and test for relative velocity biasing between Ss and Es. This by product will allow a self-consistent matching of the E and S samples into a uniform data set for future *POTENT* analysis. In §2 we compare visually the dynamical fields of velocity and mass-density as derived by *POTENT* separately from Ss and from Es. In §3 we describe the interpolation/smoothing procedure from galaxy velocities to a radial velocity field. In §4 we carry out a statistical comparison of the smoothed radial velocity fields in order to test the hypotheses. Our results are summarized and discussed in §5.

## 2. QUALITATIVE COMPARISON OF POTENT MAPS

The current sample of objects (galaxies or groups) with measured distances and redshifts has been carefully calibrated and put together from several data sets by Willick *et al.* (1993). The E data is based on the seven-samurai data as compiled by Burstein (1990), while the S data have been drastically improved since then by including Mathewson *et al.* (1992), Willick (1991), Courteau (1992), Han and Mould (1990; 1992), and Mould *et al.* (1991). The sample used here, which is a preliminary version of Willick *et al.* (1994), has 2367 spirals in 1013 objects and 546 ellipticals and S0s in 251 objects. Figure 1 compares the distribution of spirals and ellipticals in a slice of thickness $\pm 1700$ km s$^{-1}$ about the supergalactic plane. The sampling of spirals is denser and more extended, and their distance errors are smaller, so the current comparison is limited by the sampling and errors of the ellipticals. The Galactic zone of avoidance, running perpendicular to the supergalactic Y axis, is very poorly sampled in both.

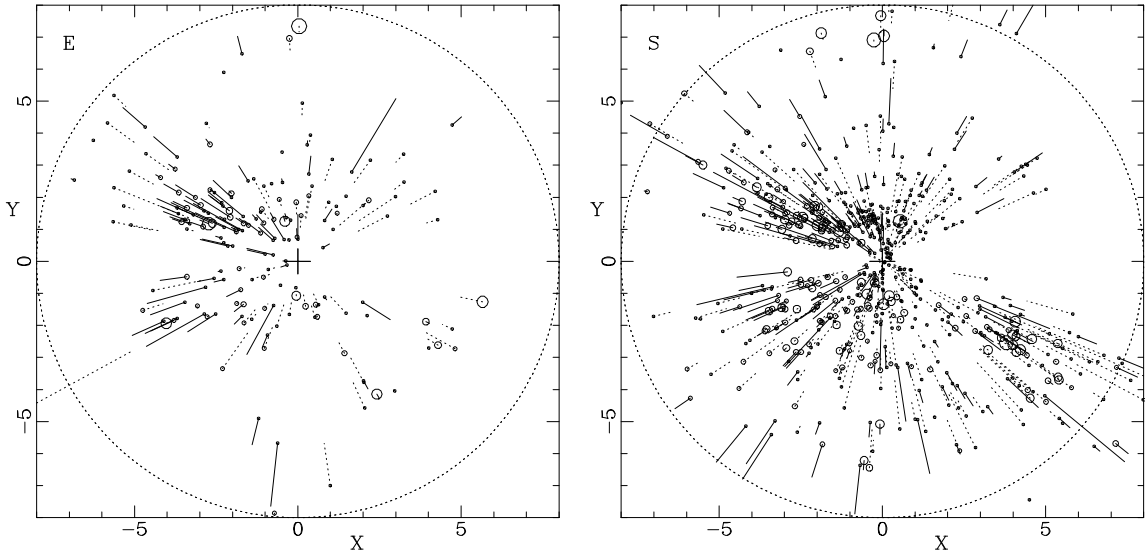

**Figure 1:** Measured radial velocities of objects projected onto the supergalactic plane in a slice of thickness $\pm 1700$ km s$^{-1}$. Distances and velocities are in units of 1000 km s$^{-1}$. The symbols mark the positions and the lines mark the peculiar velocities in the CMB frame. The solid and dashed lines help distinguishing between radial outflow and inflow. The area of each symbol is proportional to the number of galaxies making the object. The radial distances and peculiar velocities are corrected for Inhomogeneous Malmquist bias.

The figure shows the peculiar radial velocity of each object, projected on the supergalactic plane. The distances were corrected for Inhomogeneous Malmquist bias (see §3 below). The comparison of the individual velocities indicates a qualitative similarity between the velocities of the two samples in those regions that are properly covered by both, but it is naturally hard to tell how significant this similarity is.

*POTENT* takes these discrete and noisy radial velocity data, smoothes it with a tensor Gaussian window of radius 1200 km s$^{-1}$, applies the gravitational ansatz of potential flow to recover the missing tangential components of the velocity field, and finally applies a quasi-linear approximation to gravitational instability (Nusser *et al.* 1991), with an assumed value of $\Omega$, to reconstruct the associated field of mass-density fluctuations. POTENT has been applied here separately to the S data and to the E data. Figure 2 shows the recovered fields in the supergalactic plane, in comparison. The fields are shown only in regions where the reconstruction is fairly reliable in each case. The dominated limiting criterion is that the density of sampling is high enough such that the distance to the fourth neighboring galaxy, $R_4$, is less than 1500 km s$^{-1}$. We do not enforce a strong distance error criteria yet, in order to show a somewhat more extended region in the supergalactic plane than is later being used in the quantitative comparison.

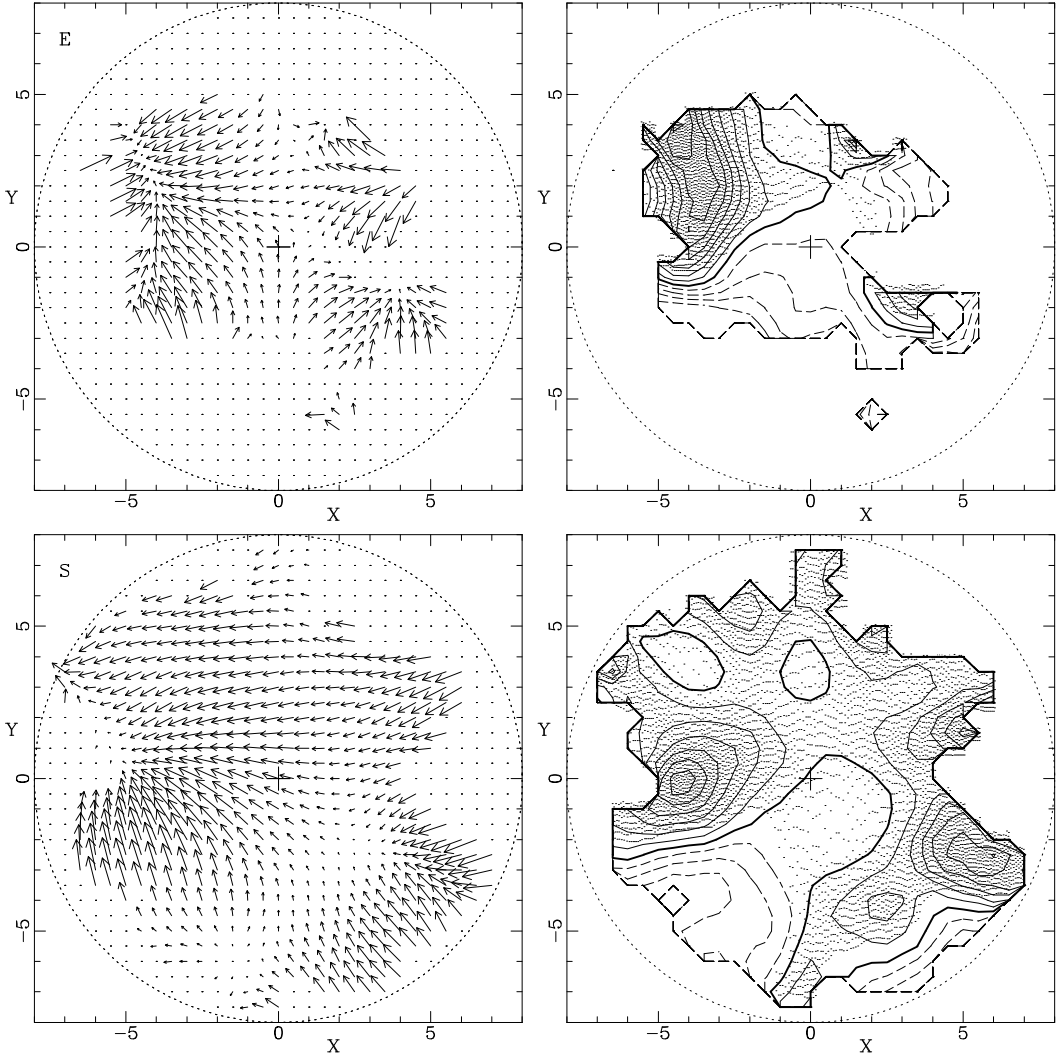

**Figure 2:** POTENT velocity and mass-density fields in the supergalactic plane as traced independently by ellipticals and S0s (top) and by spirals (bottom). Shown are only regions that are properly sampled, $R_4 < 1500$ km s$^{-1}$. Distances and velocities are in units of 1000 km s$^{-1}$. The heavy line is $\delta = 0$, solid contours mark positive fluctuations and dashed contours mark negative fluctuations, with contour spacing 0.2.

Note the small volume properly sampled by Es compared to the extended volume spanned by the Ss, in accordance with Fig. 1. The comparison is possible only within the region common to the two volumes where reliable results are obtained for Es *and* for Ss. This region is practically defined by the Es.

The immediate impression from Fig. 2 is of a general similarity between the fields of ellipticals and of spirals, both showing the same main features. A general flow from right to left, with a tendency to focus at a "Great Attractor" near $X \simeq -4000 \text{ km s}^{-1}$ (Lynden-Bell *et al.* 1988), dominates the two velocity fields. The Great Attractor is recovered in both density maps as an extended ramp of overdensity dominating the $X < 0$ hemisphere, with the Local Supercluster as an extension through Virgo ($X = 0, Y = 1000$). Both fields also recover the Perseus-Pisces supercluster at the opposite side of the sky, and the great void in between. The maps differ in their details, though. It is left for the quantitative comparison of the next section to tell whether these differences are significant under the errors.

## 3. THE SMOOTHING PROCEDURE

We carry the quantitative comparison of the E and S fields at the level of the smoothed radial velocity field; this field is recovered directly from the data without involving any assumption concerning the gravitational (irrotational) nature of the flow. Since this first part of the *POTENT* procedure is the basis for the quantitative analysis presented in the next section, we address it in some detail here.

A major issue in our analysis is the treatment of *Inhomogeneous Malmquist bias* (IM, see a review in Willick 1991; 1994). The distance errors, combined with galaxy-density variations along the line of sight, systematically enhance the inferred velocity gradients (and thus the inferred density fluctuations). This bias is reduced and corrected for by a procedure that has been tested using fake data, which were constructed from N-body simulations to mimic the observed noisy data. The procedure and its testing are discussed in detail elsewhere (Kolatt *et al.* 1994; Dekel *et al.* 1994). In brief, the galaxies are first grouped heavily in redshift space. By reducing the distance error of each group of $N$ members to $\Delta/\sqrt{N}$, where $\Delta = 0.15$ and $0.21$ for Ss and Es respectively, this grouping significantly reduces the IM bias. Then, the noisy inferred distance of each galaxy, $D$, is replaced by the expectation value of the true distance, $r$, given $D$ (Willick 1991, eq. 5.70):

$$E(r|D) = \frac{\int_0^\infty r^3 n(r) \exp\left(-\frac{[\ln(r/D)]^2}{2\Delta^2}\right) dr}{\int_0^\infty r^2 n(r) \exp\left(-\frac{[\ln(r/D)]^2}{2\Delta^2}\right) dr} . \qquad (1)$$

For single galaxies, $n(r)$ is the number density of galaxies in the underlying distribution from which galaxies were selected for the sample (by quantities that do not explicitly depend on $r$). The density run $n(r)$ is multiplied by a grouping correction factor when necessary, and is truncated at an appropriate distance in case of a redshift limit. This procedure reduces the IM bias in the fake catalog to the level of a few percents. As an approximation to $n(r)$ of the real data, we use the density of *IRAS* 1.2 Jy galaxies (Fisher *et al.* 1992), Gaussian smoothed at 500 km s$^{-1}$. The final correction typically amounts to less than 10%.

Any remaining IM bias is not expected to introduce an artificial correlation between the E and S fields. On the contrary, it's effect on the Es, which are more noisy and more clumpy, is expected to be stronger, thus weakening any true correlation.

Next, the observed radial velocities at the positions of the objects are smoothed into a radial velocity field at grid points. The aim of this procedure is to reproduce the value that would have been obtained had the true three-dimensional velocity field been sampled densely and uniformly in space, and had it been smoothed with a spherical Gaussian window of radius $R_s$. In such an ideal case, the desired velocity at each grid point is the best-fit bulk velocity for the data weighted by a three-dimensional Gaussian centered on that grid point. However, the limitations of the sampling introduce several severe biases that need to be dealt with.

1. *Tensor window function.* The radial directions from the origin to the objects do not coincide with the radial direction at the window center, so, unless the window radius is negligible compared to its distance from the origin, the radial velocities could not be averaged as scalars. Instead, we assume a parametric three-dimensional velocity model about the window center, find the most likely values of the parameters given the observed radial velocities at the objects with the window-function weights, and adopt the radial velocity of the constrained model at the center as the desired smoothed radial velocity. In the original POTENT we used a zeroth-order parametric model within each window, *i.e.* a local bulk velocity, for which the most-likely central value can be expresses in terms of a tensor window function (BD; DBF; BDFDB).

2. *Window bias.* Unfortunately, the effective weighting of the tensor window function is not spherical any more: the tensorial correction to the spherical window has a conical symmetry, with higher weights for objects that lie along radial rays that are closer to the radial ray through the window center. An example of the resulting bias is as follows. A true tangential infall pattern into the window center shows up as an artificial radial flow towards the origin, and a tangential outflow pattern shows up as a radial outflow away from the origin. This window bias is severe as long as the window radius is comparable to the distance from the origin. For $R_s = 1200$ km s$^{-1}$, out to $r = 5000$ km s$^{-1}$ from the Local Group, this bias in some regions introduces a radial infall of 300 km s$^{-1}$. A way to reduce this bias is by replacing the local bulk-velocity model with a linear velocity model. The first-order components tend to "absorb" most of the biased features of the constrained velocity model, leaving it's value at the window center a better approximation to the desired smoothed velocity. The disadvantage of a high-order model is that it picks up more efficiently undesired small-scale noise. Based on experimenting with N-body simulations we find the optimal procedure to be a first-order model fit out to $r = 4000$ km s$^{-1}$, a zeroth-order fit beyond 6000 km s$^{-1}$, and a smooth transition region in between where we adopt a weighted mean of the two kinds of fits (Kolatt *et al.* 1994; Dekel *et al.* 1994).

3. *Sampling-gradient bias.* If the true velocity field varies within the effective window, the nonuniform sampling introduces a sampling-gradient bias, which has been carefully analyzed in DBF. The averaging is galaxy weighted while we actually aim at an equal-volume weighting. To correct for this bias we weight each object by the local volume it occupies. In the original POTENT we used the inverse of the local density at the object, as estimated by the cube of the distance to its 4th neighboring object. This procedure has been improved (Kolatt *et al.* 1994; Dekel *et al.* 1994) by assigning cells from a fine grid to neighboring objects, and weighting each cell by the value of the window function at the cell position. A given object is weighted differently for windows centered at different positions. With the current sample, this procedure could reduce the sampling-gradient bias to negligible levels within the region where at least a few objects reside in the effective central region of the window. For the S sample. such regions typically extend out to $6000 \text{ km s}^{-1}$ or more from the Local Group.

4. *Random distance errors.* Unfortunately, one has to compromise again. The random distance errors become very large at large distances, where the sampling is sparse and shot noise becomes a major factor. The way to reduce the effect of noise would ideally be to weigh each object inversely by the variance of its distance measurement. This weighting, however, spoils the weighting which has been carefully designed to minimize the sampling-gradient bias. In particular, it biases the smoothed velocity towards the nearby parts of each window, where the sampling is typically denser and the distance errors are smaller, and towards nearby clusters even if they lie relatively far from the central region of the window. The compromise adopted here (as in BDFDB) is to weigh by both the volume weights and the random errors.

The smoothing procedure about a grid point $\boldsymbol{r}_0$ thus takes the following form. For a set objects at positions $\boldsymbol{r}_i$ with radial velocities $u_i$, we minimize the sum of weighted residuals

$$\chi^2 = \sum_i W_i \left[ u_i - \boldsymbol{v}(\boldsymbol{r}_i) \cdot \hat{\boldsymbol{r}}_i \right]^2 \tag{2}$$

as a function of the parameters defining the model velocity field $\boldsymbol{v}(\boldsymbol{r})$. The model velocity field is an expansion of the form

$$\boldsymbol{v}(\boldsymbol{r}) = \boldsymbol{B} + \bar{\bar{\boldsymbol{L}}} \cdot (\boldsymbol{r} - \boldsymbol{r}_0) \ . \tag{3}$$

The first term $\boldsymbol{B}$ is a zeroth-order bulk velocity, with three components. The second term is a first-order linear field, with $\bar{\bar{\boldsymbol{L}}}$ a two-dimensional symmetric tensor of six independent components, assuming an irrotational flow. The weights are

$$W_i \propto \frac{V_i}{\sigma_i} \exp[-(\boldsymbol{r}_i - \boldsymbol{r}_0)^2 / 2 R_s^2] \ , \tag{4}$$

where $V_i$ are the volume weights, $\sigma_i$ are the random distance errors, and $R_s$ is the fixed radius of the Gaussian window.

This smoothing procedure has been carefully tested using artificial samples "observed" from simulations (Dekel & Kolatt 1994; Dekel *et al.* 1994).

## 4. QUANTITATIVE COMPARISON

The comparison of the radial velocity fields of Es and Ss is done at grid points of spacing 500 km s$^{-1}$, within the joint of the volumes limited by $R_4 < 1500$ km s$^{-1}$ and $\sigma_u < 200$ km s$^{-1}$ for each of the two samples separately. The error field $\sigma_u$ is the standard deviation of the recovered *radial* velocity at a given grid point over 25 Monte-Carlo noise simulations, estimating the uncertainty due to random distance errors. Figure 3a compares point by point the radial velocity fields (hereafter $v_E$ and $v_S$ for ellipticals and spirals respectively). The errors, $\sigma_u$, are shown for a random subset of the grid points.

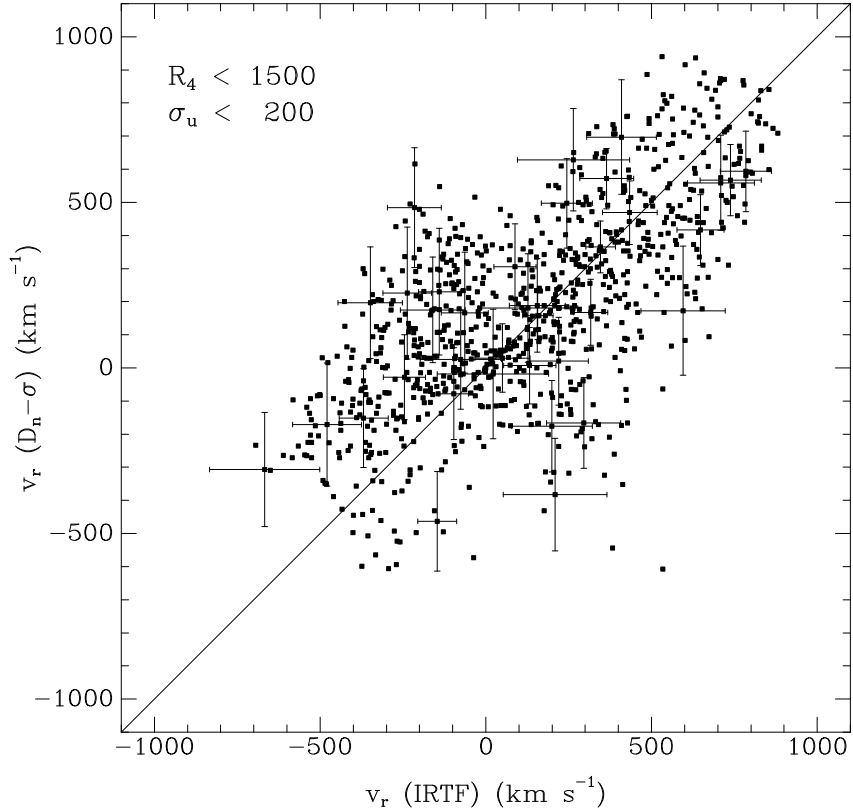

**Figure 3a:** Smoothed radial velocities as traced by spirals and by ellipticals and S0s, compared at grid points of spacing 500 km s$^{-1}$ within the comparison volume ($R_4 < 1500$ km s$^{-1}$, $\sigma_u < 200$ km s$^{-1}$). One-sigma errors, based on Monte-Carlo simulations, are shown for a random sample of points. Velocities from *The real data*.

Fig. 3a shows a clear correlation between the observed fields. To raise any doubt that this correlation is not somehow produced by the smoothing procedure from the nonuniform and noisy sample, we also sample and smooth in a similar way two completely independent radial velocity fields taken from random N-body simulations. One is sampled at the positions of the (true) Es and the other at the positions of the true Ss. These two fields are compared point by point in Figure 3b, within the same comparison volume as the real data, showing no apparent correlation.

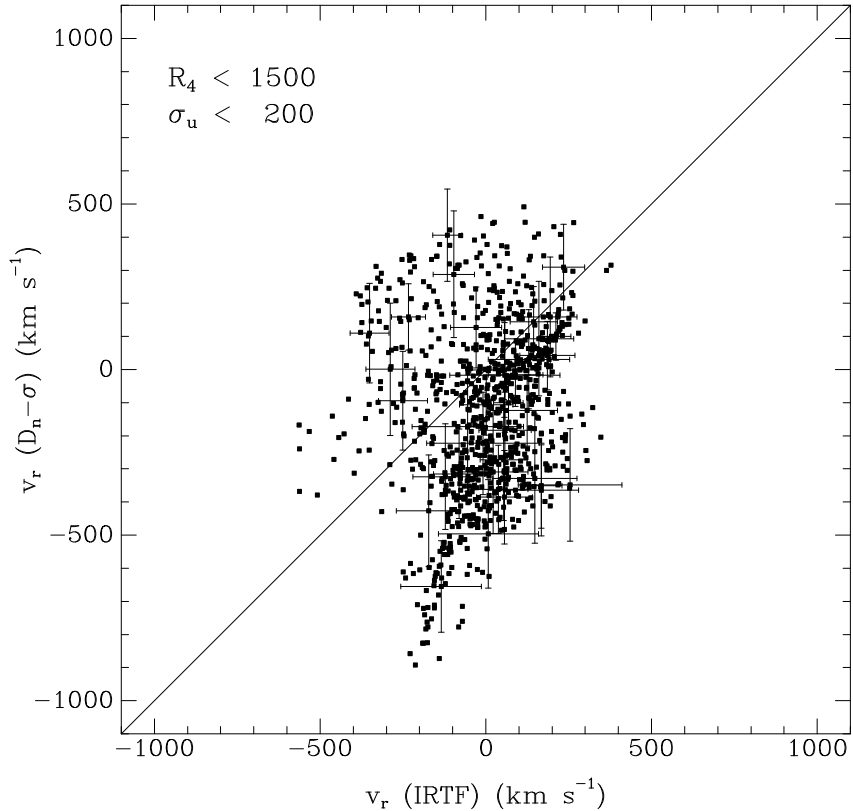

**Figure 3b:** Smoothed radial velocities as traced by spirals and by ellipticals and S0s, compared at grid points of spacing 500 km s$^{-1}$ within the comparison volume ($R_4 < 1500$ km s$^{-1}$, $\sigma_u < 200$ km s$^{-1}$). One-sigma errors, based on Monte-Carlo simulations for the real data, are shown for a random sample of points. Velocities from two *independent N-body simulations.*

To quantify the difference between two fields $v_S$ and $v_E$, with errors $\sigma_S$ and $\sigma_E$, given in a set of grid points, we define the statistic

$$D \equiv \sum \left[ \frac{(v_E - v_S)^2}{\sigma_E^2} + \frac{(v_E - v_S)^2}{\sigma_S^2} \right] \Big/ \sum \left[ \frac{(v_E + v_S)^2}{\sigma_E^2} + \frac{(v_E + v_S)^2}{\sigma_S^2} \right] , \qquad (5)$$

where the summation is over the grid points. Each term in the numerator is twice the square of the distance of a point in Fig. 3 from the diagonal line of maximum correlation, $v_E = v_S$, measured in units of the appropriate error in each direction. The denominator makes the statistic yield unity in the case of no correlation: each term here is twice the square of the distance of the point to the orthogonal diagonal, $v_E = -v_S$. Thus, $D = 0$ marks maximum correlation, $D = 1$ means no correlation, and $D > 1$ indicates anti-correlation.

There is a fundamental degree of freedom in each of the peculiar velocity fields: the estimated distances to the objects are only relative to each other, which requires that the absolute distance to one object, typically the Coma cluster, be chosen in a somewhat arbitrary way. This freedom corresponds to an arbitrary choice of zero point in each distance indicator, which allows for an arbitrary monopole term of a Hubble-like peculiar velocity component to be added to each of the peculiar velocity fields, $v \to v + \epsilon r$. There is no a priori guarantee that the preliminary choices of zero points for the spirals and for the ellipticals were done in a consistent way. For one thing, the Coma cluster as defined by the spirals is not necessarily the same as the Coma cluster defined by the ellipticals. In order to calibrate these two zero points relative to one another, we allow for an additional Hubble-type component and minimize within the comparison volume the difference $D$ between the fields $v_S$ and $v_E + \epsilon r$ as a function of the parameter $\epsilon$. The minimum is obtained at $\epsilon = 0.05$ with 27% improvement in $D$ over the preliminary $\epsilon = 0$ case. The data shown in Figure 3a is after correcting $v_E$ for the best-fit relative zero point. The resulting difference between $v_S$ and $v_E$ is $D = 0.104$.

Is this value of $D$ consistent with the hypothesis the $v_E$ and $v_S$ are both noisy versions of the same underlying velocity field? To answer this question we have to determine the probability distribution of $D$ under the distance errors using Monte-Carlo simulations, and then check how probable the observed value of $D$ is by this distribution.

Without much loss of generality, we choose the underlying field to be the 1200 km s$^{-1}$ smoothed velocity field recovered by POTENT from the full data set of all the objects together (Dekel et al. 1994). This field is sampled once at the positions of Ss and then at the positions of Es. The distance and peculiar velocity of each object is perturbed by a random Gaussian variable to mimic the intrinsic scatter in the distance indicator: a standard deviation of 15% of the distance for spirals, 21% for ellipticals, and a $\sqrt{N}$ reduction for a group of $N$ objects. We have generated 30 smoothed Monte Carlo fields for Ss and 30 for Es, and then computed $D$ for the 30x30 independently-perturbed $v_S - v_E$ pairs. Figure 4 shows the resultant distribution of the $D$ statistic under this Monte-Carlo noise procedure (histogram 1).

The observed value of $D = 0.104$ falls at the 90% percentile of the distribution: 10% of the pairs of fields, which were perturbed from the same underlying field, score worse than the data. The hypothesis of one underlying velocity field cannot be rejected by the data.

To show that this is not a trivial result of the errors being big, we also test the opposite hypothesis that the two fields are completely independent of each other. To represent independent fields we choose 30 N-body simulations of the standard CDM spectrum (Davis et al. 1985, with $\Omega = 1$, $b = 1$ and $H_0 = 50$), using a PM code (Bertschinger & Gelb 1991) with $64^3$ grid points and particles in a periodic box of side 20,000 km s$^{-1}$. The radial velocities of one simulation are sampled at the positions of the (true) spirals, and the radial velocities of another simulation are sampled at the positions of the (true) ellipticals, thus mimicking sampling biases similar to those present in the real data. The two corresponding smoothed fields yield a difference $D$. The tail of the distribution of $D$ over 450 pairs of independent fields is also shown in Fig. 4 (hashed histogram 2). The median (not shown) is near $D = 1$, as expected from the definition of $D$. Only 0.2% of the random pairs are more correlated than the observed fields. The hypothesis of no correlation between the observed fields is thus rejected with high confidence, demonstrating that the method is discriminatory despite the big errors.

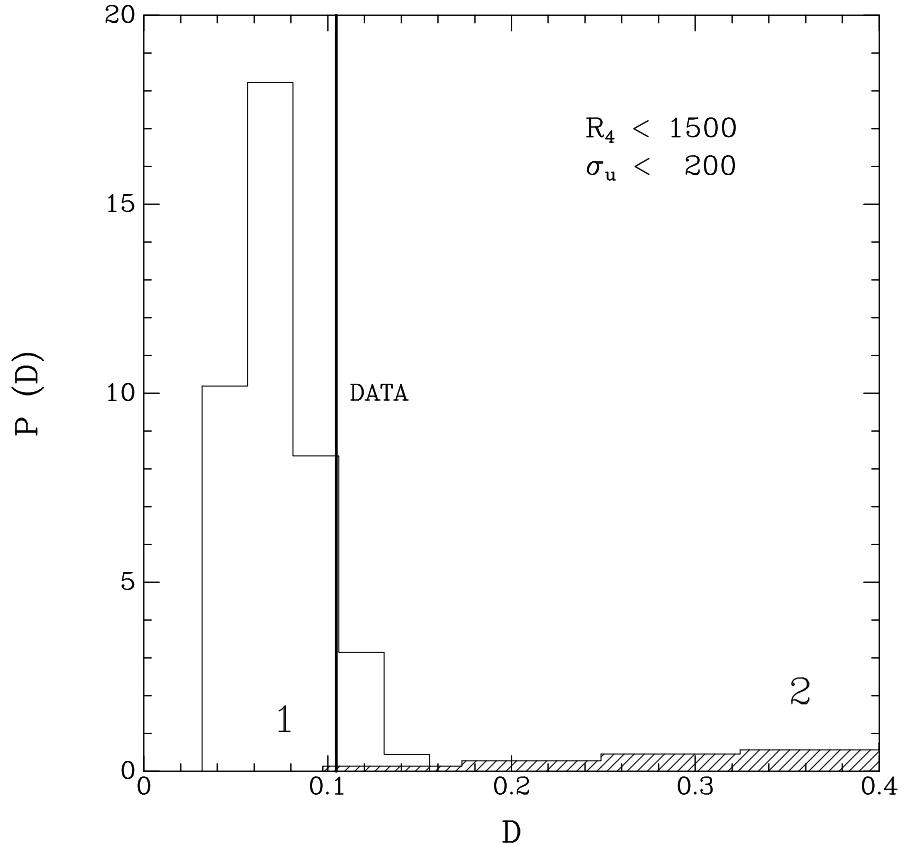

**Figure 4:** Testing the hypotheses within the comparison volume ($R_4 < 1500$ km s$^{-1}$, $\sigma_u < 200$ km s$^{-1}$). The vertical line marks the measured difference $D$ between the velocity fields traced by Ss and Es from the data. The histogram (1) is the probability distribution of the statistics $D$, derived from pairs of velocity fields which were perturbed from the same underlying field and sampled alternatively at the positions of the Ss and the Es. The hashed histogram (2) is the distribution of $D$ over pairs of independent N-body simulations, sampled at the positions of the true Ss and the Es.

The correlation between the E and S fields could also be affected by a possible type-dependent *velocity biasing*. In principle, the center-of-mass velocities of galaxies could be biased tracers of the matter velocity field because of the degree of freedom of internal velocities within galaxies, dynamical friction, or other effects. There is some evidence for a $10-20\%$ negative bias of this sort in N-body simulations with and without gas dynamics (Carlberg 1991; Carlberg & Dubinski, 1991; Gelb 1991; Katz, Hernquist & Weinberg 1992; Cen & Ostriker 1992). Since galaxies of different types have different internal dynamics and they tend to reside in different environments with different density biasing, their velocities could also be biased differently. To a first approximation, the relative velocity bias between Ss and Es could show up as a deviation from unity of the slope $v_S/v_E$ of the best-fit line minimizing the difference $D$ between $v_S$ and $v_E$. The fit yields $v_S/v_E = 1.08$, with $4\%$ improvement in $D$ relative to the unbiased case, bringing the data from the 10th to the 13th percentile of the distribution of $D$. This small improvement provides no significant evidence for a relative velocity biasing between Ss and Es.

In one more attempt to sharpen the distinction between the velocity fields of objects of different types, we repeated the above analysis for another division of the objects into two samples: "field spirals" versus "clusters" of either ellipticals or spirals. The field-cluster samples show a correlation similar to the E-S samples, with a similar relative velocity bias, $v_F/v_C = 1.08$. Thus, the field and cluster galaxies are also consistent with tracing the same underlying velocity field. The fact that the relative velocity bias is not larger in this case indicates that it is not solely a function of cluster membership.

## 5. CONCLUSIONS

Given the current noisy data, sampled in a relatively small cosmological volume, we find that the smoothed radial velocity fields derived independently from spirals and from ellipticals and S0's are *consistent* with being noisy versions of the same underlying velocity field. The opposite hypothesis that the fields are completely independent of each other is significantly ruled out by the same method, indicating that the consistency is not dominated by errors.

This result is consistent with the following set of hypotheses: (a) the peculiar velocity fields indicated by the TF and $D_n$-$\sigma$ relations are real, (b) the velocities are produced by gravity, and (c) the relative velocity biasing between spirals and ellipticals is less than 10%. The observed correlation does not prove any of these hypotheses, of course, but since this set is commonly adopted in large-scale dynamics as a working hypothesis with far-reaching implications, the consistency is encouraging.

The strength of the present test is limited by the small, sparse and noisy sample of ellipticals and S0s. Several ongoing observational efforts (e.g. EFAR by Colless *et al.* 1993) promise improvement in the size of the volume sampled. Improved techniques for measuring distances, such as Tonry's method using surface-brightness fluctuations with $\sim 5\%$ error (Tonry 1992), can drastically reduce the errors in the E sample.

Admittedly, this test cannot strictly rule out the possibility that the inferred large-scale bulk flow is mostly a reflection of a coherent environmental dependence of the distance indicators, but it makes this idea less plausible. If the environmental effects were dominant, then the E-S correlation would have implied that the two distance indicators had to conspire to have their zero points vary together in space, with a coherence length of a few tens of megaparsecs. This would have required, for example, that the large-scale properties of the environment which are responsible for these variations were different from the local properties determining the type of the galaxy. It would also have required that the properties affecting the virial equilibrium which determines the $D_n$-$\sigma$ relation, and the properties affecting the additional constraint involved in the TF relation at galaxy formation, varied together in space. It is hard to come up with plausible physical mechanisms of this sort. The one quantity that could plaussibly affect the two distance indicators in a correlated way is $M/L$. To test for this possibility one has to compare the TF and $D_n$-$\sigma$ velocities to the peculiar velocity field inferred from a completely different distance indicator, such as the one based on surface-brightness fluctuations, or on supernovae.

Another word of caution is that the observed correlation by itself does not rule out a non-gravitational origin for the velocities, such as cosmological explosions (Ostriker and Cowie 1981; Ikeuchi 1981; Ostriker, Thompson & Witten 1986), or radiation pressure instabilities (Hogan & Kaiser 1983; Hogan & White 1986). Any model in which all objects obtain the same velocities independently of their type or size would pass this test.

Nevertheless, our conclusion here is encouraging in that it adds to two other lines of evidence in support of the common working hypotheses. First, the strong correlation found between the mass density field derived from the observed peculiar velocities by *POTENT* and the galaxy density field derived from redshift samples such as the *IRAS* sample (Dekel *et al.* 1993), is consistent with the assumptions that galaxies trace mass and that gravity is responsible for the relation between the velocity and density fields (although the correlation is more sensitive to the former than to the latter assumption, see Babul *et al.* 1994). Second, the general agreement in terms of gravity between the bulk flow of $\sim 350 \, \mathrm{km\,s^{-1}}$ with a coherence length of $\sim 12,000 \, \mathrm{km\,s^{-1}}$ (*e.g.* Dekel *et al.* 1994) and the fluctuations of order $10^{-5}$ found by COBE in the Microwave Background across angular scales $> 1°$ (Smoot *et al.* 1992) is not only consistent with gravitational instability and true velocities (*e.g.* Bertschinger, Gorski & Dekel 1990); it would be rather hard to explain in a non-gravitational scenario or if the velocities were not real. The case for the reality of the inferred peculiar velocities and for the standard gravitational instability theory is thus in a pretty good shape, but the case is certainly not closed (see a review in Dekel 1994).

We thank our peculiar velocity and *POTENT* collaborators: E. Bertschinger, D. Burstein, S. Courteau, A. Dressler, S.M. Faber, J. Willick, and A. Yahil. This research has been supported by BSF grants no. 89-00194 and 92-00355, and by a Basic Research grant of the Israeli Academy no. 462/92.